\documentclass[conference]{IEEEtran}
\IEEEoverridecommandlockouts
\usepackage{cite}
\usepackage{amsmath,amssymb,amsfonts}
\usepackage{algorithmic}
\usepackage{graphicx}
\usepackage{textcomp}
\usepackage{xcolor}
\def\BibTeX{{\rm B\kern-.05em{\sc i\kern-.025em b}\kern-.08em
    T\kern-.1667em\lower.7ex\hbox{E}\kern-.125emX}}
\begin{document}

\title{Fortress: A Case Study in Stabilizing Search Recommendations via Temporal Data Augmentation and Feature Pruning\\

}

\author{\IEEEauthorblockN{1\textsuperscript{st} Milind Pandurang Jagre}
\IEEEauthorblockA{
\textit{Apple}\\
m\_jagre@apple.com}
\and
\IEEEauthorblockN{2\textsuperscript{nd} Jia Huang}
\IEEEauthorblockA{
\textit{Apple}\\
jiahuang2016@gmail.com}
\and
\IEEEauthorblockN{3\textsuperscript{rd} Dayvid V. R. Oliveira}
\IEEEauthorblockA{
\textit{Apple}\\
dvro@apple.com}
\and
\IEEEauthorblockN{4\textsuperscript{th} Zhinan Cheng}
\IEEEauthorblockA{
\textit{Apple}\\
jzcheng96@gmail.com}
\and
\IEEEauthorblockN{5\textsuperscript{th} Babak Seyed Aghazadeh}
\IEEEauthorblockA{
\textit{Apple}\\
babak9796@gmail.com}
\and
\IEEEauthorblockN{6\textsuperscript{th} Puja Das}
\IEEEauthorblockA{
\textit{Apple}\\
pdas3@apple.com}
\and
\IEEEauthorblockN{7\textsuperscript{th} Chris Alvino}
\IEEEauthorblockA{
\textit{Apple}\\
calvino@apple.com}
\and
\IEEEauthorblockN{8\textsuperscript{th} Jinda Han}
\IEEEauthorblockA{
\textit{Apple}\\
jinda\_han@apple.com}
\and
\IEEEauthorblockN{9\textsuperscript{th} Kailash Thiyagarajan}
\IEEEauthorblockA{
\textit{Apple}\\
k\_thiyagarajan@apple.com}
}

\maketitle

\begin{abstract}
In search and recommendation systems, predictive models often suffer from temporal instability when certain input features introduce volatility in output scores. This instability can degrade model reliability and user experience especially in multi-stage systems where consistent predictions are critical for downstream decision making. We introduce Fortress, a general framework for enhancing model stability and accuracy by identifying and pruning features that contribute to inconsistent prediction scores over time. Fortress leverages historical snapshots temporally partitioned datasets capturing score fluctuations for the same entity across periods and follows a four-step process: (1) collect historical snapshots, (2) identify samples with unstable predictions, (3) isolate and remove instability- inducing features, and (4) retrain models using only stable features. While semantic features from LLMs and BERT-based models improve generalization, they often lack full query or entity coverage. Engagement-based features offer strong predictive power but tend to introduce temporal instability. Fortress mitigates this trade-off by suppressing the volatility of engagement signals while retaining their predictive value leading to more stable and accurate models. We validate Fortress on a query-to-app relevance model in a large-scale app marketplace. Offline experiments demonstrate notable improvements in prediction stability (measured by Coefficient of Variation) and classification performance (measured by PR-AUC).
\end{abstract}

\section{Introduction}
Recommendation systems must provide not only relevant results, but also consistent and predictable ones. Instability in model predictions where the same query-app pair yields significantly different scores across time can lead to poor user experience and undermine trust in the system. While recent advancements in LLM-based features have improved contextual relevance [2, 4, 10], engineered features derived from organic user engagement behavior (clicks, searches, etc.) continue to play a crucial role [7, 8, 12, 14, 15]. However, these features can introduce temporal inconsistencies that affect score stability. 

Model instability often arises when certain features cause the output score to vary across temporal snapshots, leading to inconsistent user experiences. In search-based recommendation systems, these score fluctuations can cause items to alternate between being shown and not shown in search results. Such instability can affect the overall quality of the recommendations, leading to a less consistent and predictable user experience.

In this work, we present Fortress, a generalizable framework for improving model prediction stability through feature selection by leveraging historical snapshots of model inputs and outputs over time. The core insight is that some features though potentially predictive in isolated snapshots may introduce volatility across time. By identifying and pruning such features, we can enhance both model robustness and accuracy. This method fits into the broader class of “wrapper” feature selection methods, where a feature subset is selected based on validation performance, as introduced in [6]. A review of feature selection methods can be found in [11].

We evaluate this method on a relevance prediction task for an app marketplace, where the goal is to classify query-app pairs as relevant or not. While this case study grounds the methodology, the technique is broadly applicable to any scenario where a scoring model evaluates the same entity over time, such as user-item recommendations. Our results show improved prediction consistency as measured by decreased Coefficient of Variation (CV) [5] and improved classification performance as measured by increased PR-AUC.

\section{THE FORTRESS FRAMEWORK}
Our method assumes that certain features may lead to unstable predictions when used in model scoring across time. We define a feature as unstable if its inclusion causes prediction volatility across historical snapshots for the same entity. Historical snapshots are temporally partitioned datasets contiguous intervals over a time period allowing us to evaluate prediction consistency for entities scored multiple times.

\begin{figure}[htbp]
\centering 
\includegraphics[width=\columnwidth]{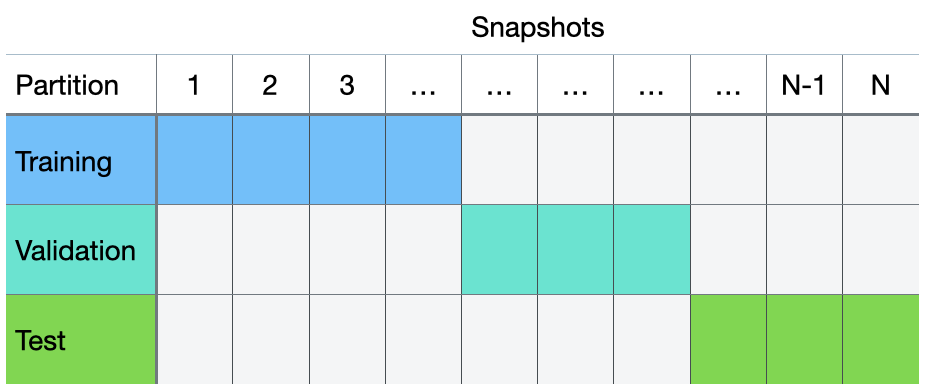} 
\caption{Representation of multi-snapshot approach with data sampled across different historical points, and disjoint samples between training, validation, and test sets.}
\label{fig}
\end{figure}

We demonstrate the method using a dataset of query-app pairs from a major commercial app marketplace. Human raters label each pair as BAD, ACCEPTABLE, GOOD, or EXCELLENT. To simplify evaluation, we treat this as a classification task where the model learns to filter out BAD pairs. 

We utilized a quarter of the data, consisting of daily snapshots, for training and validation purposes. The entire dataset was divided into training, test, and validation datasets, following the industry standard of 70\%, 15\%, and 15\% respectively. The entire dataset consisted of approximately 40\% pairs labeled as BAD, while the remaining 60\% were divided equally among pairs labeled as ACCEPTABLE, GOOD, or EXCELLENT.

Our evaluation metrics include PR-AUC to assess classification performance, and Coefficient of Variation (CV) to measure score stability, defined as standard deviation divided by mean for prediction scores across snapshots.

The proposed method proceeds as follows:
\begin{enumerate}
    \item Obtain multiple snapshots over time. Each entity (e.g., query-app pair) and its snapshot data is assigned to a single data partition.
    \item Identify unstable samples with high score variance across snapshots.
    \item Identify and remove features that contribute to instability while offering limited performance gain.
    \item Retrain the model using the reduced feature set and all snapshots.
\end{enumerate}

To determine which features to remove, we first identify samples corresponding to the 75th percentile highest CV across temporal snapshots. For these high-variance samples, we compute the CV for each feature individually. Features with the highest CVs are considered candidates for pruning. However, we only proceed with feature removal if doing so results in statistically significant improved PR-AUC on the validation set as compared to baseline. We employ a greedy search approach to compute the CV by iteratively removing features deemed for pruning during model retraining without those features. This process continues until the pruned feature list is exhausted, and we achieve a statistically significant PR-AUC lift. 

This is particularly important in systems with multi-stage retrieval and ranking pipelines, where instability in prediction scores can propagate through downstream stages and cause significant shifts in overall system behavior.

\section{EXPERIMENTS AND RESULTS}
We evaluated our method using an XGBoost classifier [1] to predict an app relevance in the context of a query, by training on largely, two categories of features, (a) Semantic Representation (SR): LLM-based and DistilBERT based features and (b) Engagement features.

We used two types of SR features. The first is an LLM-based [13] feature designed to capture query-to-app similarity independent of engagement signals. For this, the system employs a proprietary Foundation Model [3], which leverages query and app metadata to generate relevance scores. The second is based on a DistilBERT model [9] in a two-tower architecture trained with triplet loss, where the model learns to embed queries and items such that relevant pairs are closer and irrelevant pairs are pushed further apart in the embedding space.

The engagement features are non-personalized and based on highly aggregated historical user interactions with the query-app pair (clicks, searches etc.). Our model consisted of around 25 features, which were refreshed several times daily and are the main source of predicted score instability.

XGBoost was selected for its speed and robustness to noise. As is typical in classification modeling, we interpret the output score as a probability and use that score in our recommendation system where stability is desirable.

Table I summarizes results for the following:



\begin{table}[h]
\centering
\caption{ACCURACY AND STABILITY METRICS}
\label{table_metrics}
\resizebox{\columnwidth}{!}{
\begin{tabular}{|l|c|c|}
\hline
\textbf{Method} & \textbf{PR-AUC (95\% CI)} & \textbf{CV (95\% CI)} \\ \hline
SR features (LLM + DistilBERT) only & 0.9367$\pm$0.0028 & 0.4744$\pm$0.0057 \\ \hline
All features (SR + engagement) & 0.9572$\pm$0.0023 & 0.5462$\pm$0.0056 \\ \hline
Multi-snapshot with all features & 0.9593$\pm$0.0022 & 0.5283$\pm$0.0056 \\ \hline
Fortress & 0.9595$\pm$0.0022 & 0.5274$\pm$0.0056 \\ \hline
\end{tabular}%
}
\end{table}

The results show that while adding engagement features improves PR-AUC, it also increases score volatility. Multi-snapshot training mitigates some of that instability. Our proposed method, with both multi-snapshot and removal of unstable features, improves score volatility by 3.44\% and classification PR-AUC by 0.24

As a downstream effect of the improved prediction score stability, Table II presents the relative improvement of Fortress compared to the single-snapshot approach. In particular, it shows the relative reduction in the number of (query, app) pairs that exhibit flip-flop instability across time that is, pairs that appear in the final stage of our recommendation system in one historical snapshot but are absent in another. This metric reflects the system-level impact of more stable predictions on the consistency of recommendations over time. The results show a consistent improvement across regions, along with a healthy global gain in stability for our recommendation system. Here, 'regions' refer to groups of countries; Exact country groups are proprietary and have been masked.

\begin{table}[htbp]
\caption{Relative Reduction in Candidate Flip-Flop Rate by Region}
\begin{center}
\begin{tabular}{|l|c|}
\hline
\textbf{Region} & \textbf{Relative Reduction in Flip-Flop Rate} \\
\hline
Region 1 & 14.7\% \\
\hline
Region 2 & 17.0\% \\
\hline
Region 3 & 29.2\% \\
\hline
Region 4 & 49.6\% \\
\hline
Region 5 & 28.0\% \\
\hline
Region 6 & 38.2\% \\
\hline
\hline
\textbf{Global} & \textbf{25.8\%} \\
\hline
\end{tabular}
\label{tab_flipflop}
\end{center}
\end{table}

Interestingly, some features that appeared beneficial in a single snapshot were found to degrade performance when evaluated across time. This reinforces the value of temporal multi-snapshot analysis for robust feature selection.

Semantic features, for example LLM-based features tend to be more stable across time, as content embeddings and app metadata change slowly unless retrained. However, relying solely on LLM-based features can overlook key signals from user behavior. These features are also dependent on the underlying dataset snapshot on which the model was trained and can sometimes be limited in their knowledge or coverage (approx. 80\%, for our study). In contrast, engagement-based features evolve with real-time user interactions and can fill these gaps over time. Our method offers a principled way to incorporate engagement signals while managing their volatility.

\section{CONCLUSION}
Delivering a high-quality user experience requires recommendation systems to produce results that are both relevant and consistent. While engagement-based features enhance model accuracy, they can introduce prediction instability over time.

We proposed Fortress, a general framework for improving model stability by performing temporal data augmentation and pruning features that contribute to score inconsistency, while preserving overall prediction accuracy. Using historical snapshots, we identified prediction-volatility-inducing features and retrained models without them. This method improved both model stability and classification performance in our case study and is applicable to a broad range of recommendation systems where scoring consistency matters. In multistage retrieval and ranking systems, volatility in production scores can lead to undesirable shifts in system dynamics and user experience resulting in suboptimal outcomes. Our approach applies to such systems, where maintaining scoring stability is critical to ensuring reliability and performance.



\end{document}